\documentstyle[aps,manuscript]{revtex}

\newcommand{\be}{\begin{equation}}
\newcommand{\ee}{\end{equation}}

\def\fun#1#2{\lower3.6pt\vbox{\baselineskip0pt\lineskip.9pt
\ialign{$\mathsurround=0pt#1\hfil##\hfil$\crcr#2\crcr\sim\crcr}}}
\begin{document}

\def\baselinestretch{1.5}
\normalsize

\title{On the thermal mass shift of nucleons}
\author{V. L. Eletsky$^{1,2}$ and B. L. Ioffe$^{1}$}
\address{$^1$Institute of Theoretical
and Experimental Physics,
 B.Cheremushkinskaya 25, Moscow 117259, Russia}
 \address{$^2$Institut f\"ur Theoretische Physik III, Universit\"at
Erlangen-N\"urnberg, D-91058 Erlangen, Germany}

\maketitle
\begin{abstract}
The mass shift $\Delta m_N$ of a nucleon in the finite temperature pion gas 
is calculated in the case of an arbitrary nucleon momentum. A general formula 
is used which relates the in-medium
mass shift $\Delta m(E)$ of a particle to the real part of the forward
scattering amplitude ${\rm Re} f(E)$ of this particle on constituents of
the medium and
its applicability domain is formulated.  The mass shift of the nucleon
in thermal equilibrium with pion gas is also calculated.

\end{abstract}
\newpage
\def\baselinestretch{1.5}
\normalsize
\vspace{2mm}
\vspace{2mm}
The problem of how the properties of hadrons change in hadron or nuclear matter
in comparison to their free values has attracted a lot of attention
recently. Among these properties of immediate interest are the in-medium mass
shifts of particles. Different models as well as model independent approaches
were used to calculate hadron mass shifts both at finite temperature and finite
density (for a review, see e.g. \cite{1}). It is however clear on physical grounds
that the in-medium mass shift of a particle is only due to its interaction with
the constituents of the medium. Thus one can use phenomenological information on
this interaction to calculate the mass shifts. In our recent paper\cite{2} we have argued
that the mass shift of a particle in medium at relatively low density of the latter
can be related to the forward scattering amplitude $f(E)$ of this particle on the constituents
of the medium

\be
\Delta m (E) = -2\pi\frac{\rho}{m}{\rm Re} f(E)
\label{dm0}
\ee
Here $m$ is the vacuum mass of the particle, $E$ is its energy in the rest frame of the
constituent particle, and $\rho$ is the density of consituents. The normalization of the
amplitude corresponds to the standard form of the optical theorem

\be
k\sigma = 4\pi {\rm Im} f(E),
\label{opt}
\ee
where $k$ is the particle momentum. All quantities in Eqs.(1) and (2) 
correspond to the rest frame of the target.
In most of the papers on the in-medium hadron mass shifts the hadrons were considered at
rest.
As seen from Eq.(1) this restriction is not necessary theoretically.
Experimentally it is also desirable to have theoretical predictions in broad
energy interval, since it extends the possibilities of experimental
investigations. As discussed in \cite{2} for the cases of $\rho$ or
$\pi$-mesons embedded in nuclear matter the energy dependence of the mass
shifts is rather significant at low energies, i.e. in the resonance region.
Here we will demonstrate this for the case of the thermal mass
shift of a nucleon imbedded in a pion gas.

For the nucleon at rest the mass shift was calculated by Leutwyler and 
Smilga\cite{3} by
using experimental information\cite{4} on the $\pi N$ forward scattering 
amplitude. We will
show how their resut follows from Eq.(\ref{dm0}) and then generalize 
it to the case of a
moving nucleon. First, let us go from the rest frame of the nucleon to 
that of a pion. Under this transformation
$\rho\to\rho\sqrt{1-v^2}$, ($v$ is the pion velocity).
Kinematically, we have $m_NE_{\pi}=m_{\pi}E_N$, where $E_{\pi}$ is the pion
energy in the nucleon rest frame, $E_N$ is the nucleon energy in the pion
rest frame. Since the cross section $\sigma$  is the same in both the rest
frames, we have from the optical  theorem $f_{N\pi}(E_N)/E_N=f_{\pi
N}(E_{\pi})/E_{\pi}$, where the indeces $N\pi$ and $\pi N$ of the amplitudes mean pion and
nucleon rest frames correspondingly. Then Eq.(\ref{dm0}) can be rewritten as

\be
\Delta m_N (E=m_N) = -2\pi\frac{\rho}{E_{\pi}}{\rm Re} f_{\pi N}(E_{\pi})
\label{dm1}
\ee
Here $\rho$ is now the pion gas density in the rest frame of the nucleon and
$f_{\pi N}(E_{\pi})$ is the amplitude of forward scattering of a pion with energy $E_{\pi}$
on the nucleon in this frame. Since the rest frame of the nucleon is also
the rest frame of the pion gas, we recover the formula used in \cite{3}

\be
\Delta m_N  = -2\pi\sum_{\pi^a} \int\frac{d^3 k}{(2\pi )^3}
\frac{1}{E_{\pi}}\frac{1}{e^{E_{\pi}/T}-1}{\rm Re} f_{\pi^a N}(E_{\pi})
\label{dmsl}
\ee
where $E_{\pi}=\sqrt{k^2 - m_{\pi}^2}$. The simplest way to generalize this formula
to the case of a nucleon with a finite momentum moving through a pion gas which is at rest
is to go to the nucleon rest frame again boosting the pion gas. Since the pion phase space
$d^3 k/E_{\pi}$ is Lorentz invariant we only need to boost the pion distribution function
which is achieved by substituting $E_{\pi}$ for $p_{\pi}u$ in the Bose factor, where
$u=(1/\sqrt{1-v^2}, {\bf v}/\sqrt{1-v^2})$ is the 4-velocity of the nucleon. Using
${\bf v}={\bf p_N}/E_N$ and $\sqrt{1-v^2}=m_N/E_N$ and performing the integral over the
angle between ${\bf p}_{\pi}$ and ${\bf p}_N$ we finally get

\be
\Delta m_N  = -T\frac{m_N}{p_N}\sum_{\pi^a}\int\limits^{\infty}_{m_{\pi}}
\frac{dE_{\pi}}{2\pi}\log\frac{1-e^{-E_{+}(E_{\pi})/T}}{1-e^{-E_{-}(E_{\pi})/T}}
f_{\pi^a N}(E_{\pi})
\label{dmp}
\ee
where $E_{\pm}=(E_N E_{\pi}\pm p_N p_{\pi})/m_N$.

We have calculated $\Delta m_N (T)$ for $p_N=0.5, 1$ and $5$ GeV using the same data\cite{4}
on the isospin symmetric forward amplitude
\be
D^{+}(E_{\pi})=\frac{4\pi}{3}\sum_{\pi^a} f_{\pi^a N}(E_{\pi})
\label{d}
\ee
as in\cite{2}. The results are shown in Fig.1 and compared to the case of the nucleon at
rest\cite{2}. As pointed out in\cite{3} the dip in $\Delta m_N (T)$ for the nucleon at rest
at low $T$ is due to
the $\Delta$-resonance. Indeed, the amplitude $D^{+}$ is positive in the corresponding
resonance energy. It becomes negative at higher energies, but their contribution to the
mass shift is exponentially suppressed by the statistical factor in Eq.(\ref{dmsl}).
However, if the nucleon has a finite momentum then in its rest frame the scattering of pions is
shifted to higher energies. At energies above the $\Delta$-resonance, $D^{+}$ is mainly negative.
Thus, it could be expected that at sufficiently high nucleon momentum the effect of the
of $\Delta$ would be washed out, and the dip will disappear. As is seen from Fig.1 the dip
decreases and indeed disappears at $p_N\sim 1$ GeV.

We have also calculated $\Delta m_N (T)$ for the case when the nucleon is in thermal
equilibrium with the heat bath.
Strictly speaking in this case we should also average over Maxwell
distribution of nucleon momenta. For simplicity instead we put nucleon
energy fixed and equal to the mean energy in the pion gas, $E_N = m_N +(3/2)T$ . 
The results of calculations  are presented in Fig.2.
One can see that the original dip at $T\approx 90$ MeV has practically
disappeared:  it has moved to $T\approx 70$ MeV and its depth decreased by a
factor of two.

The effective nucleon width (damping rate) in the thermal pion gas is obtained
in a similar way 

\be
\gamma = T\frac{m_N}{p_N}\sum_{\pi^a}\int\limits^{\infty}_{m_{\pi}}
\frac{d E_{\pi}(E_{\pi}^2 -m_{\pi}^2)^{1/2}}{(2\pi)^2}
\log\frac{1-e^{-E_+(E_{\pi})/T}}{1-e^{-E_-(E_{\pi})/T}}\sigma_{\pi^{a} N}(E_{\pi})
\label{g}
\ee
The effect of a finite momentum of the nucleon on $\gamma$ is much less than on
the mass shift, since there is no sign change in the integrand of Eq.(\ref{g})
contrary to the case of Eq.(\ref{dmp}).
In the case of nucleon in thermal equilibrium with pion gas we have
numerically at $T=100$ MeV, $\gamma =33$ MeV. For the nucleon at rest at this
temperature $\gamma =30$ MeV\cite{3}.

Let us estimate the applicability domain of our approach. The main
restriction arises from the requirement (see \cite{2}) 
$|{\rm Re}f_{N\pi}(E_N)|<d$,
where $d$ is the mean distance between pions in the gas. 
We use the relation
$f_{N\pi}(E_N)=(E_N/E_{\pi})f_{\pi N}(E_{\pi})$ and note that
in the whole important energy interval
$|{\rm Re}f_{\pi N}| < 0.7$ fm (apart from the region of 
$\Delta$-resonance, where $-{\rm Re} f_{\pi^{+}N}$ reaches $1.5$ fm). 
At $T=150$MeV the kinematical factor $E_N/E_{\pi}\approx 2.6$
for thermal pions, while $d\approx 2$ fm and rapidly increases at lower $T$. 
We thus see that this approach works up to about $T=150$ MeV, where the 
approximation of non-interacting pion gas also becomes invalid. 
Another restriction, $\lambda_{N}=1/p_{N}\ll d$ which could be violated at
low $T$ is in fact fulfilled for a thermalized nucleon due to the exponential
growth of $d$ at $T\to 0$.

This work was supported in part by INTAS Grant 93-0283, CRDF grant RP2-132, 
RFFR grant 97-02-16131 and
Schweizerischer National Fonds grant 7SUPJ048716 V. L. E. acknowledges
support of BMBF, Bonn, Germany.

\begin{figure}
\caption{Temperature dependence of $\Delta m_N$ for $p_N =0$, 0.5, 1, and 1,5 GeV
(curves $a, b, c$ and $d$, respectively).}
\end{figure}
\begin{figure}
\caption{Temperature dependence of $\Delta m_N$ for a thermalized nucleon,
$E_N =m_N +(3/2)T$, (curve $b$) in comparison with the one for the nucleon at rest (curve $a$).}
\end{figure}
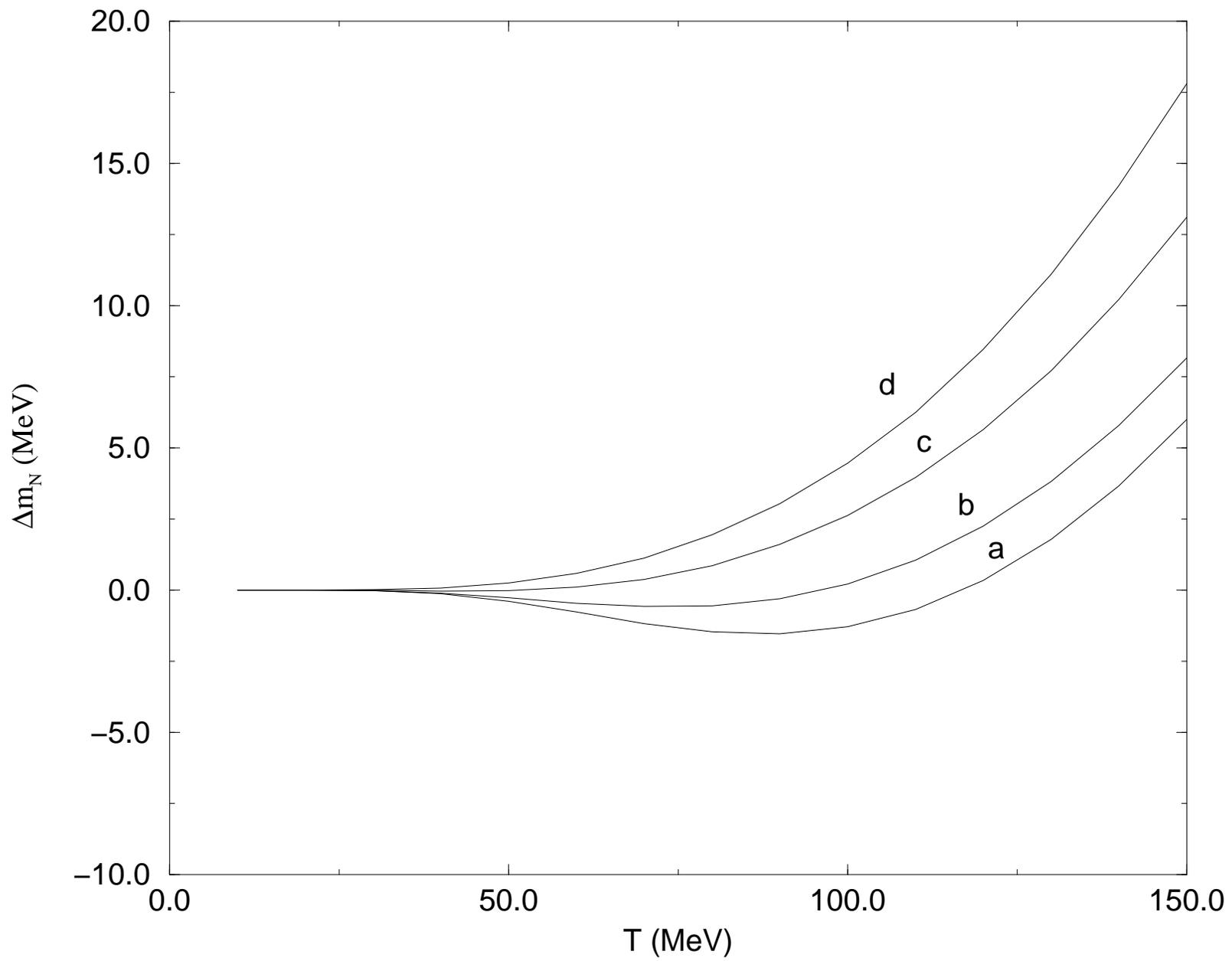
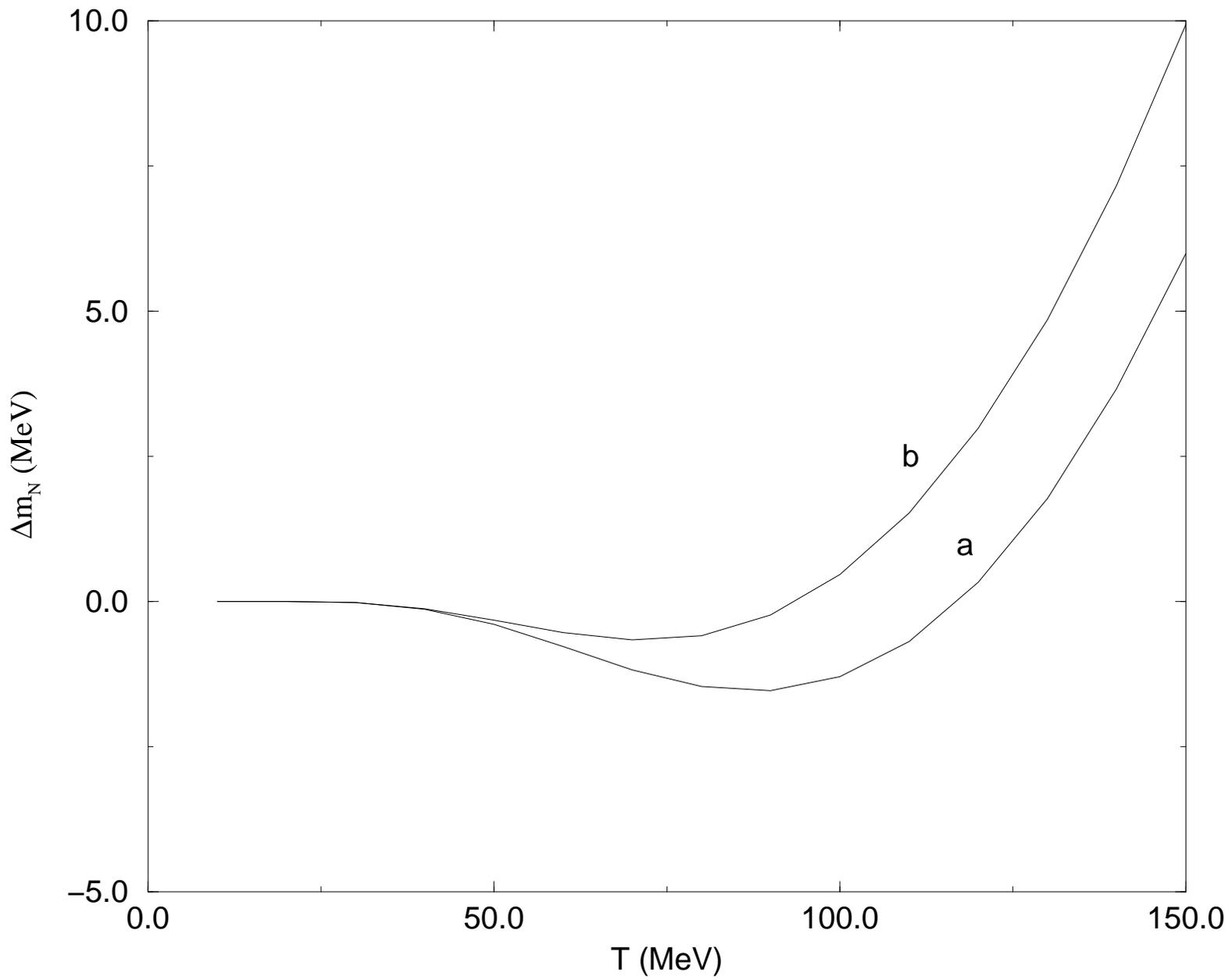
\end{document}